\documentclass[aps,prd,floats,floatfix,amssymb,twocolumn,superscriptaddress,nofootinbib]{revtex4-1}

\usepackage{amsmath}
\usepackage{amsfonts}
\usepackage{mathrsfs}
\usepackage{graphicx,graphics}
\usepackage{hyperref}
\usepackage{booktabs}

\DeclareMathOperator{\sech}{sech}

\DeclareMathOperator{\arccoth}{arcoth} 

\usepackage{verbatim}
\usepackage{lipsum}

\usepackage{color}

\definecolor{gray}{rgb}{0.5, 0.5, 0.5}
\definecolor{babyblue}{rgb}{0.54, 0.81, 0.94}
\definecolor{babyblueeyes}{rgb}{0.63, 0.79, 0.95}
\definecolor{palebrown}{rgb}{0.6, 0.46, 0.33}

\newcommand{\rhat}{\hat{r}_{*}}

\newcommand{\psileft}{\psi^{\scriptscriptstyle (<)}_\omega}
\newcommand{\psiright}{\psi^{\scriptscriptstyle (>)}_\omega}
\newcommand{\hleft}{h^{\scriptscriptstyle (<)}}
\newcommand{\hright}{h^{\scriptscriptstyle (>)}}

\begin{document}
	
	\title{Robin boundary conditions in acoustic BTZ black holes}
	
	\author{Christyan C. de Oliveira}
	\email{chris@ifi.unicamp.br}
	\affiliation{Instituto de F\'isica ``Gleb Wataghin'', Universidade Estadual de Campinas, 13083-859, Campinas, SP, Brazil}
	\author{Ricardo A. Mosna}
	\email{mosna@unicamp.br}
	\affiliation{Departamento de Matem\'atica Aplicada, Universidade Estadual de Campinas, 13083-859, Campinas, SP, Brazil}
	\author{Jo\~ao Paulo M. Pitelli}
	\email{pitelli@unicamp.br}
	\affiliation{Departamento de Matem\'atica Aplicada, Universidade Estadual de Campinas, 13083-859, Campinas, SP, Brazil}
	
	\begin{abstract}
		We introduce an analog model for the conformally coupled scalar field on the BTZ black hole. The model is based on the propagation of acoustic waves in a Laval nozzle. Since the BTZ black hole is not a globally hyperbolic spacetime, the dynamics of the scalar field is not well defined until extra boundary conditions are prescribed at its spatial infinity.  We show that quasinormal modes (QNMs) satisfying Dirichlet, Neumann, and Robin boundary conditions in the BTZ black hole can be interpreted in terms of ordinary QNMs defined with respect to an appropriately extended nozzle. We also discuss the stability of our model with respect to small perturbations. 
	\end{abstract}
	
	\maketitle
	
	\section{Introduction}
	\label{sec:intro}
	It is well known that acoustic waves propagating in an inviscid fluid mimic scalar waves in a black hole background. This was first shown by Unruh in 1981 \cite{unruh1981experimental} and, since then, physicists have found many other systems exhibiting this same feature (see \cite{barcelo2011analogue} for an extensive list). In particular, the past decades have witnessed the emergence of an increasing number of such \textit{analog gravity models} designed to probe a variety of phenomena in black hole backgrounds. For instance, a number of experiments have been performed to observe the analog of rotational superradiance \cite{torres2017rotational}, quasinormal ringing \cite{torres2019analogue,torres2020quasinormal} and Hawking radiation \cite{rousseaux2008observation,weinfurtner2011measurement,euve2016observation,steinhauer2016observation,de2019observation,kolobov2021observation}.
	
	Aside from experimental tests, analog models are also important for theoretical purposes, since they provide new insights for many kinds of problems in different geometries. A particularly interesting class of geometries is that given by {\it nonglobally hyperbolic} spacetimes. In these spaces, a deterministic time evolution is not well defined until extra boundary conditions are prescribed \cite{wald1980dynamics,ishibashi2003dynamics,ishibashi2004dynamics}. This occurs because information can reach (or come from) spatial infinity in finite time. Hence, information coming from infinity  can influence the spacetime bulk dynamics. Ilustrative examples where analog models are used to provide theoretical insights about nonglobally hyperbolic spacetimes were given in Refs.  \cite{mosna2016analogue, de2022analogue}.

	A particularly interesting example of nonglobally hyperbolic space is the Ba\~{n}ados, Teitelboim and Zanelli (BTZ) black hole \cite{banados1992black}, which is an asymptotically anti-de Sitter solution for three-dimensional general relativity (GR) with a negative cosmological constant \cite{carlip19952+}. For this spacetime, the effective potential describing scalar wave propagation does not vanish at spatial infinity so that the solution does not behave like a plane wave there. Hence, the usual outgoing boundary condition used to define quasinormal modes (QNMs) in asymptotically flat spacetimes cannot be fulfilled \cite{birmingham1997three,myung2003difference,dappiaggi2018superradiance}. Thus, for the BTZ black hole, QNMs will depend not only on the effective potential, but also on the boundary condition at infinity.
	
	In this work, we introduce an analog model for the conformally coupled scalar field on a BTZ black hole based on a Laval nozzle, which is a convergent-divergent nozzle with a throat in the middle, usually employed to accelerate air \cite{roshko1957elements}. By establishing a sufficiently strong difference of pressure between the nozzle ends, a transonic flow regime can be achieved. On one side of the nozzle, there is a subsonic flow; on the other side, a supersonic flow is established. The sonic point (where air velocity equals sound velocity) is located at the nozzle throat.

	We find that the obtained nozzle has a finite length, with its end corresponding to the spatial infinity of the BTZ spacetime, so that our analog model effectively maps the exterior region of the BTZ black hole into a finite region of the Laval nozzle.
	Since the effective potential governing the wave propagation does not vanish at the nozzle end (which corresponds to the BTZ spatial infinity), we still cannot impose plane wave outgoing boundary conditions to find QNMs. In order to circumvent this problem, we consider a family of extensions for the nozzle. We choose the extensions in such a way that the corresponding effective potentials go to zero in the new end. By doing this, we recover the plane wave behavior and we can thus impose outgoing boundary conditions and find the ordinary QNMs of the extended nozzle (corresponding to BTZ spacetime + extension). In this way, we interpret the QNMs of the conformally coupled scalar field on the BTZ black hole (which do not obey outgoing boundary conditions at spatial infinity) in terms of ordinary QNMs of acoustic waves in the nozzle (which do obey the usual outgoing boundary conditions). We find that the ordinary QNMs can be sorted according to their parity and show that odd ordinary QNMs correspond to QNMs in the black hole which satisfy a Dirichlet boundary condition; and even ordinary QNMs correspond to black hole QNMs obeying Neumann or Robin boundary conditions.
	
	Finally, we use a result from the dynamics of the scalar field in the BTZ black hole \cite{dappiaggi2018superradiance} to discuss the stability of our model under linear perturbations.
	
	This paper is organized as follows. In Sec. \ref{sec:The nozzle analogue to the BTZ black hole}, we briefly review the equations of acoustics in the Laval nozzle, and apply the method of \cite{abdalla2007perturbations} to find the nozzle for which acoustic waves correspond to those of a conformally coupled field on the BTZ black hole. In Sec. \ref{sec:Robin boundary conditions in the BTZ analogue nozzle}, we consider continuations of the effective potential of Sec. \ref{sec:The nozzle analogue to the BTZ black hole} to find extensions for the nozzle previously obtained. We also show how one may use the ordinary QNMs of acoustic waves to emulate QNMs obeying Dirichlet, Neumann and Robin boundary conditions at BTZ spatial infinity.  After that, we discuss the stability of our model under small perturbations. Finally, we discuss our results in Sec. \ref{sec:conclusion}.

	\section{The nozzle analog to the BTZ black hole}
	\label{sec:The nozzle analogue to the BTZ black hole}
	\subsection{Wave propagation in the BTZ black hole}
	The scalar field $\Psi$ conformally coupled to the BTZ geometry is governed by the equation \cite{lee1998probing,garbarz2017scalar}
	\begin{align}
	\left(\Box + \frac{3}{4 l^{2}} \right)\Psi =0,
	\label{eq:eq of motion conformal}
	\end{align}
	where the d'Alembertian operator, $\Box = \nabla_{\mu}\nabla^{\mu}$, is calculated with respect to the spacetime metric	
	\begin{align}
	ds^{2} = -\left( -M +\frac{r^{2}}{l^{2}}  \right)dt^{2} + \left(-M + \frac{r^{2}}{l^{2}}\right)^{-1} dr^{2} + r^{2} d\phi^{2}.
	\end{align}
	Here, $M$ stands for the black hole mass (which is dimensionless for this spacetime) and $l$ is the radius of the associated three-dimensional anti-de Sitter space (AdS$_{3}$).
	
	Separating variables by 
	\begin{align}
	\Psi(t,r,\phi) = \frac{\psi(r)}{r^{1/2}} e^{-i \omega t }e^{i m \phi},
	\end{align}
	the equation of motion \eqref{eq:eq of motion conformal} yields
	\begin{align}
	-\frac{d^{2}\psi(r_{*})}{dr^{2}_{*}} + V_{\mbox{{\scriptsize BTZ}}}(r_{*}) \psi(r_*) = \omega^{2} \psi(r_{*}),
	\label{eq:Schrodinger BTZ}
	\end{align}
	where the tortoise coordinate is taken as 
	\begin{align}
	r_{*} =- \frac{l}{\sqrt{M}} \arccoth \left( \frac{r}{l\sqrt{M} } \right),
	\end{align}
	and the effective potential is given by
	\begin{align}
	V_{\mbox{{\scriptsize BTZ}}}(r_{*}) = \frac{(4 m^{2} + M)}{4 l^{2}} \sech^{2}\left( \frac{\sqrt{M} r_{*}}{l} \right).
	\label{eq:effective potential BTZ}
	\end{align}
	
	The black hole horizon $r = l\sqrt{M}$ corresponds to $r_{*}=-\infty$ and the conformal boundary at $r=\infty$ is mapped to $r_{*}=0$. For later convenience, we rescale $r_{*}$ and $\omega$ to the dimensionless quantities $\hat{r}_{*} = (\sqrt{M}/l) r_{*}$ and $\hat{\omega} = (l/\sqrt{M}) \omega$. The radial equation of motion, Eq. \eqref{eq:Schrodinger BTZ}, then becomes 
	\begin{align}
	-\frac{d^{2}\psi(\hat{r}_{*})}{d\hat{r}^{2}_{*}} + \hat{V}_{\mbox{{\scriptsize BTZ}}}(\hat{r}_{*}) \psi(\hat{r}_*) = \hat{\omega}^{2} \psi(\hat{r}_{*}),
	\label{eq:Schrodinger BTZ nondimensional}
	\end{align}
	with the dimensionless effective potential  
	\begin{align}
	\hat{V}_{\mbox{{\scriptsize BTZ}}}(\hat{r}_{*}) = \left(\frac{4 m^{2} +M}{4M}\right) \sech^{2} \hat{r}_{*} .
	\label{eq:effective potential BTZ nondimensional}
	\end{align}
	
	We intend to simulate the scalar field propagation determined by the effective potential $	\hat{V}_{\mbox{{\scriptsize BTZ}}}(\hat{r}_{*})$ in terms of acoustic waves propagating in an appropriately designed Laval nozzle. In order to achieve this, we need to know how the shape of the nozzle determines the wave propagation. In what follows, we review the fundamental equations of fluid dynamics in the Laval nozzle and show how the cross-sectional area determines the effective potential for acoustic waves.

	\subsection{Wave propagation in the Laval nozzle}
	Let us take the $x$ coordinate along the axial direction of the nozzle. We consider air as a perfect fluid flowing in a quasi-one-dimensional regime, where physical quantities vary along the $x$ axis only. The equations of motion are then the continuity and Euler's equations,
	\begin{align}
	\partial_{t} \left(\rho A \right)+ \partial_{x} \left( \rho v A \right) =0 , \label{eq:continuity equation} \\
	\rho \left( \partial_{t} + v \partial_{x}\right)v =-\partial_{x} p, \label{eq:Euler equation}
	%\partial_{t} \left(\rho v A \right) + \partial_{x} \left[ (\rho v^{2} +p)A \right]=0, \\
	%\partial_{t} \left(\epsilon  A \right) + \partial_{x} \left[ (\epsilon +p) vA \right]=0,	
	\end{align}
	where $\rho$ is the air density, $p$ is the pressure, $v$ is the air velocity and $A$ is the nozzle cross-sectional area. Furthermore, we shall assume the gas is isentropic
	\begin{align}
	p \propto \rho ^{\gamma},
	\label{eq:isentropic flow}
	\end{align}
	where $\gamma=7/5$ stands for the heat capacity ratio of air.
	
	Assuming an irrotational flow, $v=\partial_{x} \Phi$, and defining the specific enthalpy $h(\rho) = \int \rho^{-1} dp $, Eq.~\eqref{eq:Euler equation} can be reduced to the Bernoulli's equation
	\begin{align}
	\partial_{t} \Phi + \frac{1}{2} \left( \partial_{x} \Phi \right)^{2} + h(\rho) =0.
	\end{align} 
	
	To derive the linearized wave equation for sound, we first rewrite $(\rho, \Phi)$ as the sum of a contribution corresponding to the background flow $(\bar{\rho}, \bar{\Phi} )$ and a contribution corresponding to the acoustic disturbance $(\delta \rho, \phi )$. The background and perturbation satisfy 
	\begin{equation}
	\begin{array}{llccc}
	\rho= & \bar{\rho} +\delta \rho, & &  & \bar{\rho} \gg |\delta \rho|, \\
	\Phi= & \bar{\Phi} + \phi,&  &  &|\partial_{x}\bar{\Phi}| \gg |\partial_{x}\phi|. \\    
	\end{array}
	\end{equation}
	Following \cite{okuzumi2007quasinormal}, we define the auxiliary quantities 
	\begin{align}
	g =& \frac{\bar{\rho} A}{c_{s}} = \frac{\bar{\rho} A}{\sqrt{\gamma \bar{p}/\bar{\rho}}}, \label{eq:g definition}\\
	f(x) =& \int \frac{|v|dx}{c_{s}^{2} - v^{2}}, \\
	H_{\omega} =& g^{1/2} \int dt e^{i\omega[t-f(x)]} \phi(t,x),\\
	x_{*} =& c_{s0} \int \frac{dx}{c_{s} (1-\mathcal{M}^{2})}, \label{eq:x tortoise}
	\end{align} 
	where $c_{s}= \sqrt{\partial p/\partial \rho}=\sqrt{\gamma \bar{p}/\bar{\rho}}$ is the local sound speed, $c_{s0}$ is the stagnation sound speed, constant over the isentropic flow, and $\mathcal{M}=|v|/c_{s}$ is the Mach number. In terms of these quantities, the wave equation reduces to 
	\begin{equation}
	-\frac{d^{2}H_{\omega}}{dx_{*}^{2}}  + V(x_{*}) H_{\omega} = \kappa^{2} H_{\omega}, 
	\label{eq:Schrodinger Laval}
	\end{equation}
	where 
	\begin{equation}
	\kappa=\frac{\omega}{c_{s0}},
	\end{equation}
	and the effective potential is given by
	\begin{align}
	V(x_{*}) = \frac{1}{g^{2}} \left[ \frac{g}{2} \frac{d^{2}g}{dx_{*}^{2}} - \frac{1}{4}\left( \frac{dg}{dx_{*}} \right)^{2} \right].
	\label{eq:effective potential Laval}
	\end{align}
	
	The effective potential $V(x_{*})$ characterizes the dynamics of acoustic waves in the gas flow. For a transonic flow in a Laval nozzle, all the nondimensional quantities $(\rho/\rho_{0}$, $p/p_{0}$, $\mathcal{M}, \dots)$ are uniquely determined by the function $A(x_{*})/A_{th}$, where $A_{th}$ is the cross-sectional area at the throat of the nozzle \cite{roshko1957elements,okuzumi2007quasinormal}. In particular, the function $g(x_{*})$ and the effective potential $V(x_{*})$ are also completely determined by $A(x_{*})/A_{th}$.  
	On the other hand $A$ (and hence all other physical quantities) can be fully determined in terms of $g$ by the equations relating the physical variables in the nozzle. 
	
	Let us see more closely how one can express the physical quantities in terms of $g$. First, we note that it follows from Eqs. \eqref{eq:isentropic flow} and \eqref{eq:g definition} that 
	\begin{align}
	g \propto \frac{\bar{\rho} A}{\bar{\rho}^{(\gamma -1)/2}},
	\label{eq:g prov}
	\end{align}
	and from \cite{roshko1957elements} we have
	\begin{align}
	\left(\frac{A}{A_{th}}\right)^{-1}  = \frac{1}{\eta_{\gamma}}\left[1- \left(\frac{\bar{\rho}}{\rho_0}\right)^{(\gamma -1)} \right] ^{1/2} \frac{\bar{\rho}}{\rho_0},
	\label{eq:A prov} 
	\end{align}
	where $\rho_0$ is the stagnation density and 
	\begin{equation}
	\eta_{\gamma} = \sqrt{\frac{\gamma -1}{2}} \left(\frac{2}{\gamma +1}\right)^{\frac{\gamma +1}{2(\gamma -1)}}.
	\end{equation}
	
	Since Eq. \eqref{eq:Schrodinger Laval} is invariant under rescalings of $g$, we take the coefficient in Eq. \eqref{eq:g prov} so that
	\begin{align}
	g       =  \frac{\frac{A}{\eta_{\gamma}A_{th}}\frac{\bar{\rho}}{\rho_0} }{2 \left(\frac{\bar{\rho}}{\rho_0}\right)^{(\gamma -1)/2}}.
	\end{align}

	With the assumptions above, we can implement the same reasoning of \cite{abdalla2007perturbations} to find the physical variables in terms of $g$: 
	\begin{align}
	\frac{A}{A_{th}} & = \frac{\eta_{\gamma} \sqrt{2}\left[ 2 g^{2} \left(1- \sqrt{1-g^{-2}}\right) \right]^{1/(\gamma -1)}  }{\sqrt{1-\sqrt{1-g^{-2}}}},
	\label{eq:A de g} \\
	\left(\frac{\bar{\rho}}{\rho_0}\right)^{1-\gamma} & = 2g^{2} \left( 1- \sqrt{1-g^{-2}} \right), \label{eq:rho de g}\\
	c_s  &= \frac{c_{s0}}{\sqrt{2 g^{2} \left( 1- \sqrt{1-g^{-2}} \right)}}, \label{eq:cs de g}\\
	\mathcal{M}^{2} & = \frac{2}{\gamma -1} \left( 2 g^{2} \left( 1- \sqrt{1-g^{-2}} \right) -1 \right).
	\label{eq:Mach number}
	\end{align}

	For convenience, we rescale $x_{*}$ and $\omega$ to dimensionless quantities $\hat{x}_{*} $ and $\hat{\omega}$ such that
	\begin{align}
	\hat{x}_{*} = \frac{x_{*}}{L},
	\end{align}
	and
	\begin{align}
	\kappa = \frac{\omega}{c_{s0}} = \frac{\hat{\omega}}{c_{s0} T} = \frac{\hat{\omega}}{L},
	\end{align}
	where $L$ is a characteristic length in the nozzle and a characteristic time interval was chosen as $T = L/c_{s0} $. Equation \eqref{eq:Schrodinger Laval} then yields
	\begin{equation}
	-\frac{d^{2}H_{\omega}}{d\hat{x}_{*}^{2}}  + \hat{V}(\hat{x}_{*}) H_{\omega} = \hat{\omega}^{2} H_{\omega},
	\label{eq:Schrodinger Laval nondimensional}
	\end{equation}
	where the dimensionless effective potential is given by
	\begin{align}
	\hat{V}(\hat{x}_{*}) = \frac{1}{g^{2}} \left[ \frac{g}{2} \frac{d^{2}g}{d\hat{x}_{*}^{2}} - \frac{1}{4}\left( \frac{dg}{d\hat{x}_{*}} \right)^{2} \right].
	\label{eq:effective potential Laval nondimensional}
	\end{align}
	
	\subsection{Inverse problem}
	The calculations above show how the nozzle shape, given by $A(x_{*})$, determines the wave propagation in the nozzle by means of the effective potential $\hat{V}(x_{*})$. We now want to find a nozzle shape which mimics the effective potential $\hat{V}_{\mbox{{\scriptsize BTZ}}}$ for perturbations in the BTZ black hole background.
	
	As mentioned before, all physical quantities describing the flow in the Laval nozzle can be determined from the cross section $A$. On the other hand, given an effective potential, say $\hat{V}=\hat{V}_{\mbox{{\scriptsize BTZ}}}$, we should be able to find $g$ by solving Eq.~ \eqref{eq:effective potential Laval nondimensional}. This in turn determines the shape of the nozzle by means of Eq.~(\ref{eq:A de g}). A boundary condition for $g$ is given by imposing that the air and sound velocities are equal at the acoustic horizon, $|v|=c_s$, so that, from Eq.~(\ref{eq:Mach number}),
	\begin{align}
	g|_{\mbox{{\scriptsize horizon}}} = \frac{\gamma +1}{2 \sqrt{2} \sqrt{\gamma -1}}=\frac{3}{\sqrt{5}}.
	\label{eq:g horizon}
	\end{align}
	
	Before equating $\hat{V}_{\mbox{{\scriptsize BTZ}}}$ to the effective potential in the nozzle, we need to relate the radial coordinate $r$ of the BTZ spacetime to the coordinate along the nozzle $x$. In order to achieve this, we identify the respective tortoise coordinates, $d\hat{r}_{*} = d\hat{x}_{*} $. From Eqs. \eqref{eq:x tortoise}, \eqref{eq:cs de g} and \eqref{eq:Mach number}, we have 
	\begin{align}
	d\hat{r}_{*}=d\hat{x}_{*} &= \frac{c_{s0} d\hat{x}}{c_s (1-\mathcal{M}^{2})} \nonumber\\
	&= \frac{\sqrt{2g^{2} \left( 1- \sqrt{1- g^{-2}} \right)}d\hat{x}}{1 - \frac{2}{\gamma -1}\left[ 2 g^{2} \left( 1-\sqrt{1-g^{-2}} \right) -1 \right]},
	\end{align}  
	so that
	\begin{align}
	\frac{d \hat{x}}{d\hat{r}_{*}}= \frac{1 - \frac{2}{\gamma -1}\left[ 2 g^{2} \left( 1-\sqrt{1-g^{-2}} \right) -1 \right]}{\sqrt{2g^{2} \left( 1- \sqrt{1- g^{-2}} \right)}},
	\label{eq:dxdrstar}
	\end{align} 
	where $\hat{x}$ is the nondimensional coordinate related to the coordinate along the nozzle by $\hat{x} = x/L$.
	
	Since the tortoise coordinates for the nozzle and for the BTZ spacetime were made identical, we can now equate the effective potentials, Eqs. \eqref{eq:effective potential Laval nondimensional} and \eqref{eq:effective potential BTZ nondimensional}, to obtain
	\begin{align}
	\frac{g''(\hat{r}_{*})}{2 g(\hat{r}_{*})}-\frac{g'(\hat{r}_{*})^2}{4 g(\hat{r}_{*})^2}=\hat{V}_{\mbox{{\scriptsize BTZ}}}(\hat{r}_{*}),
	\label{eq:g equation}
	\end{align}
	where the prime indicates differentiation with respect to $\hat{r}_{*}$, $g' = dg/d\hat{r}_*$. This equation can be simplified by the substitution \cite{cuyubamba2013laval}
	\begin{align}
	g(\hat{r}_{*}) =  h^{2}(\hat{r}_{*})
	\label{eq:g de h}
	\end{align}
	that yields
	\begin{align}
	- h''(\hat{r}_{*}) + \hat{V}_{\mbox{{\scriptsize BTZ}}}(\hat{r}_{*})h(\hat{r}_{*}) = 0.
	\label{eq:h equation}
	\end{align}
	
	To obtain the configuration of the nozzle corresponding to the effective potential in Eq. \eqref{eq:effective potential BTZ nondimensional}, we have to solve Eq. \eqref{eq:h equation} and use Eq. \eqref{eq:g de h} and the boundary condition Eq. \eqref{eq:g horizon} to determine $g$. Then, by Eqs. \eqref{eq:A de g} and \eqref{eq:dxdrstar} we can find the cross section $A$ as a function of the coordinate $x$ along the nozzle, such that the sound propagation now mimics a scalar field propagating on the BTZ spacetime.
	
	\subsection{The Laval nozzle for the conformal scalar field propagating on the BTZ black hole}
	To keep the calculations as simple as possible, we are going to consider the mode solution with angular momentum $m=0$ (the case $m \ne 0$ can be treated in a similar fashion and is briefly discussed in Sec.~\ref{sec:conclusion}). For this case, the general solution to the differential equation \eqref{eq:h equation} can be expressed as a linear combination\footnote{Hereafter, we will drop the hat in $\rhat$ to keep the notation simpler.}
	\begin{align}
	h(r_{*}) =  c_{1} h_{1}(r_{*}) + c_{2} h_{2}(r_{*}),
	\end{align}
	where $c_{1}$, $c_{2}$, are constants and 
	\begin{align}
	h_{1}(r_{*}) & = P_{-\frac{1}{2} }(\tanh r_*), \label{eq:h1}\\
	h_{2}(r_{*}) & = Q_{-\frac{1}{2}}(\tanh r_{*}). \label{eq:h2}
	\end{align}
	Here $P_{\mu}(z)$ and $Q_{\mu}(z)$ stand for the Legendre functions of the first and second kinds, respectively \cite{olver2010nist}.
	
	The boundary condition Eq. \eqref{eq:g horizon} requires $h(r_{*})$ to be finite at the acoustic horizon ($r \to - \infty$). Since $h_{1}(r_{*})$ goes to infinity at $r \to -\infty$, we should take $c_{1} = 0$ and $c_2$ so that
	\begin{align}
	h(r_{*}) = \frac{2}{\pi}\sqrt{\frac{3}{\sqrt{5}}} \, Q_{-\frac{1}{2}}(\tanh r_{*}). 
	\label{eq:h btz}
	\end{align}
	
\begin{figure}[th!]
		\centering
		\includegraphics[width=\columnwidth]{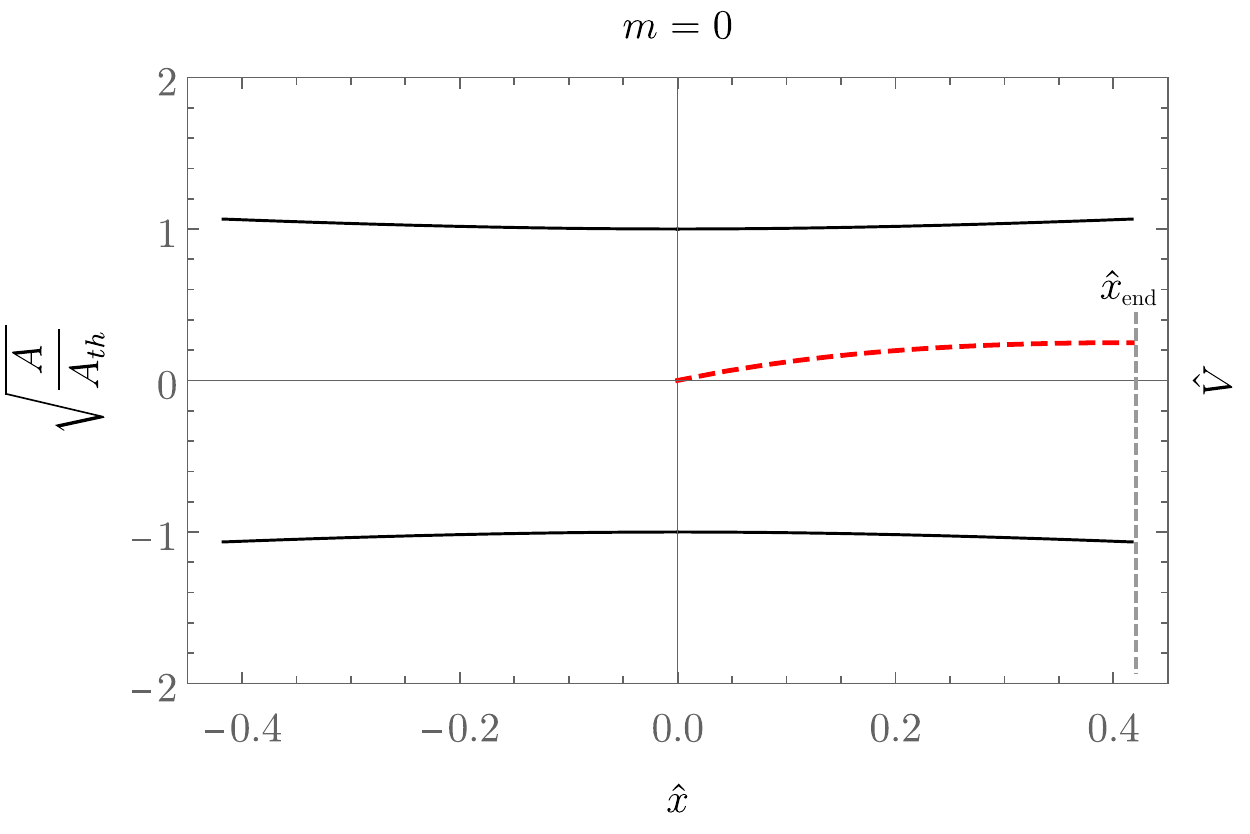}
		\caption{(Black curve) Lateral section of the Laval nozzle corresponding to the conformally coupled scalar field. The region $\hat{x}>0$ ($\hat{x}<0$) corresponds to subsonic (supersonic) flow. The sonic point (where the fluid velocity equals the sound velocity) is located at the throat $\hat{x}=0$. (Red dashed line) Nondimensional effective potential for acoustic waves in the subsonic region.}
		\label{fig:tubo BTZ}
	\end{figure}

	Having obtained $h(r_{*})$, we use Eqs. \eqref{eq:g de h}, \eqref{eq:A de g} and \eqref{eq:dxdrstar} to find the cross section $A$ as a function of the coordinate along the nozzle.     The lateral section of the resulting nozzle is represented by the black solid curve in Fig.~\ref{fig:tubo BTZ}, where we also plotted the effective potential (red dashed curve).  
	The exterior region of the BTZ black hole corresponds to the subsonic region ($x>0$).
	
	\begin{figure}[th!]
		\centering
		\includegraphics[width=\columnwidth]{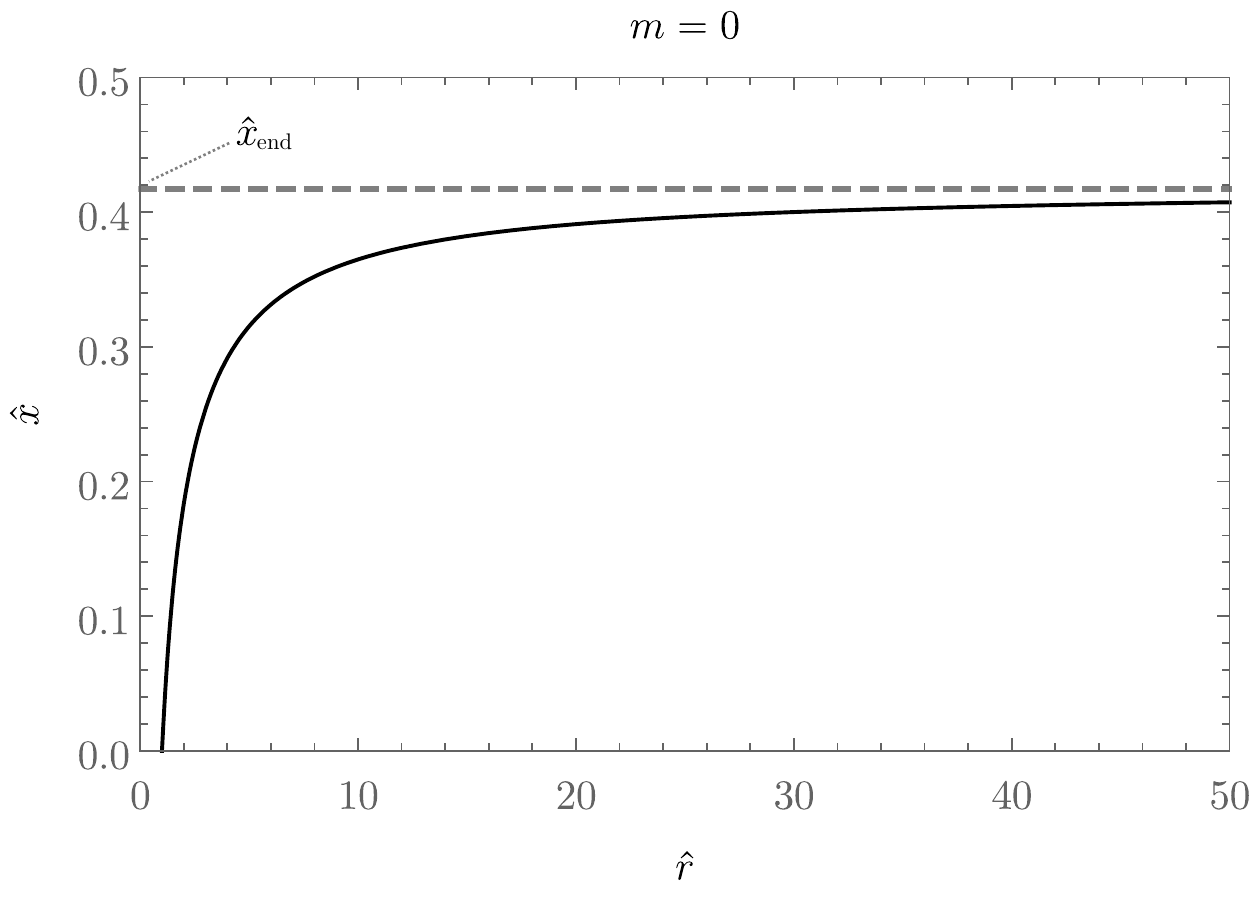}
		\caption{The nondimensional coordinate along the nozzle $\hat{x}$ as a function of the nondimensional radial coordinate $\hat{r}$ of the BTZ spacetime. The coordinate $\hat{x}$ has a finite upper limit, $\hat{x}_{{\scriptsize \mbox{end}}} \cong 0.417306$, which means that the corresponding nozzle has a finite length. The upper limit in the coordinate $\hat{x}$ is mapped into the spatial infinity of the BTZ spacetime.}
		\label{fig:tubo BTZ x de r}
	\end{figure}
	Figure \ref{fig:tubo BTZ x de r} shows the relation between the $x$ coordinate along the nozzle  and the radial coordinate $r$ of the BTZ black hole.	We observe that $x$ has a finite upper limit, at $x_{\text{end}}\cong 0.417306$. Hence the obtained nozzle has a finite length, with the upper limit of $x$ being mapped into the spatial infinity of the BTZ black hole. In other words, this means that the exterior region of the BTZ black hole is mapped into a finite region in the laboratory, with the spatial infinity of the BTZ spacetime being mapped into the right end of the nozzle.
	
	It follows that, in order to completely determine the acoustic wave propagation in the nozzle, it is necessary to prescribe a boundary condition at its right end. At the BTZ spacetime level, the necessity for a boundary condition at spatial infinity comes from its lack of global hyperbolicity \cite{garbarz2017scalar}, \cite{wald1980dynamics,ishibashi2003dynamics,ishibashi2004dynamics}. Therefore, via the correspondence found above, our model simulates the needed boundary conditions at the conformal boundary of the BTZ spacetime by appropriate boundary conditions at the nozzle (finite) right end.
		
	The boundary conditions that are compatible with sensible dynamics for the scalar field propagating in the BTZ spacetime were studied in \cite{garbarz2017scalar}. In particular, for the conformally coupled scalar field, Garbaz {\it et al.} found that Robin boundary  conditions (RBCs),
	\begin{align}
	\frac{d \psi/dr_{*}}{\psi}\bigg|_{r_{*}=0} = \beta  ,
	\label{eq:robin bc}
	\end{align}
	lead to an unambiguous time evolution. In this case, $\beta = \pm \infty$ corresponds to the Dirichlet boundary condition at infinity, $\psi|_{r_{*}=0} = 0$, and $\beta = 0$ corresponds to the Neumann boundary condition, $d \psi/dr_{*}|_{r_{*}=0} = 0$. Aside from that, in \cite{dappiaggi2018superradiance}, Dappiaggi {\it et al.} calculated the effect of RBCs on the quasinormal modes of the scalar field in the BTZ black hole. In the next section, we propose a nozzle configuration appropriate to realize QNMs obeying RBCs in the BTZ black hole.

	\section{Robin boundary conditions in the BTZ analog nozzle}
	\label{sec:Robin boundary conditions in the BTZ analogue nozzle}
	
	\subsection{Nozzle extension}
	\label{sub:nozzle extension}
	
	We have seen above that the nozzle which mimics the BTZ spacetime would abruptly end at a finite distance from the throat, at $x=x_{\text{end}}\cong 0.417306$. In what follows we continue the nozzle in such a way that the usual boundary condition for QNMs at its far right, $x\to\infty$, induces RBCs at $x_{\text{end}}$. In the $r_{*}$ coordinate, this corresponds to extending the potential $V_{\mbox{{\scriptsize BTZ}}}(r_{*})$ to the region $r_{*}>0$ (recall that the original range of the coordinate $r_{*}$ is from $-\infty$ to $0$).
	
\begin{figure*}[th!]
		\centering
		\includegraphics[width=1.8\columnwidth]{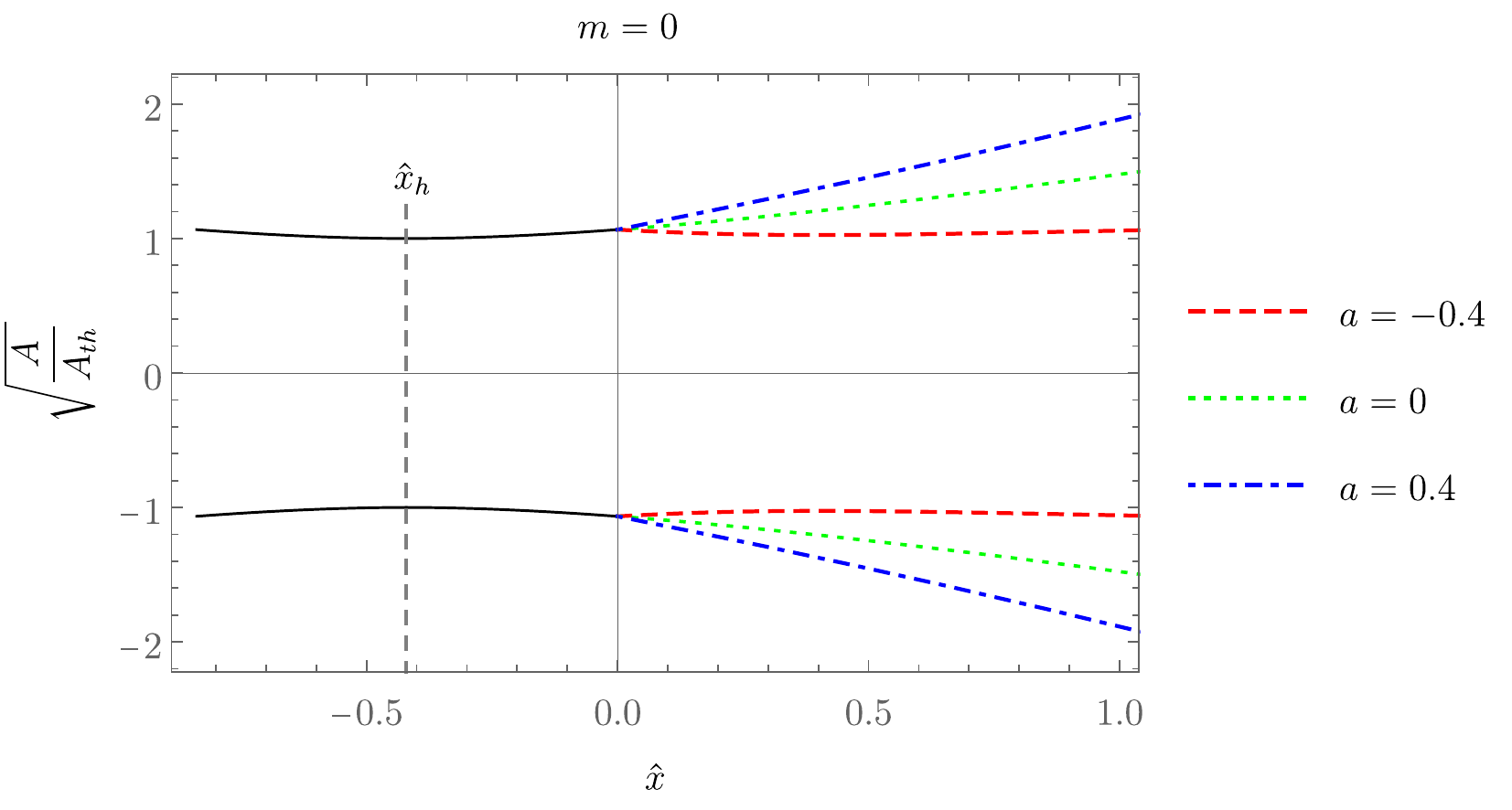}
		\caption{Lateral section of the extended Laval nozzle with different values of parameter $a$. Each value determines a different extension for the effective potential. Since the wave phenomena is mainly determined by the effective potential, different values of $a$ will lead to different quasinormal spectra. Notice that we have translated the $\hat{x}$ axis by $\hat{x} \to  \hat{x} - \hat{x}_{\mbox{{\scriptsize end}}}$ so that now the origin $\hat{x}=0$ corresponds to BTZ spatial infinity and the horizon corresponds to $\hat{x}_{h} = - \hat{x}_{\mbox{{\scriptsize end}}}\cong- 0.417306$.}
		\label{fig:extensions}
	\end{figure*}

	We will consider the following extension of $V_{\mbox{{\scriptsize BTZ}}}(r_{*})$ for $r_{*}\ge0$:
	\begin{align}
	V_{\mbox{{\scriptsize eff}}}(r_{*}) %& = V_{\mbox{{\scriptsize BTZ}}}(r_{*}) + a \, \delta(r_{*}) \nonumber \\
	&= \left[\frac{4 m^{2} +M}{M}\right] \sech^{2} r_{*} + a \, \delta(r_{*}), 
	\label{eq:effective potential BTZ extended}
	\end{align}
	where $\delta(r_{*})$ is the Dirac delta function, $a$ is a constant, and $-\infty< r_{*} < \infty $ . 
	We note that, for $-\infty < r_{*}<0$, this effective potential reduces to Eq. \eqref{eq:effective potential BTZ nondimensional}. Moreover, for $r_{*} \to \infty$, $V_{\mbox{{\scriptsize eff}}} $ goes to zero and we recover the plane wave behavior, typical for asymptotically flat spacetimes, for the field (i.e., $\psi \sim e^{\pm i \omega r_{*}}$, when $r_{*} \to \infty$). In particular, this implies that usual outgoing boundary conditions for QNMs can now be imposed in the extended model. The delta function term has the effect of producing a shape change in $A(x)$ at $x=0$ (see Fig.~\ref{fig:extensions}), which will be explored in what follows to implement the RBCs in the BTZ spacetime.
	
	Let us calculate the shape of the extended nozzle which corresponds to the extended potential above. We do this by solving Eq. \eqref{eq:h equation} with $V(r_{*})$ given by Eq. \eqref{eq:effective potential BTZ extended}.
	
	For $r_{*}<0$, the calculations are identical to the case treated in the previous section. Thus, the corresponding solution is given by Eq. \eqref{eq:h btz}. For convenience, we now denote this solution by $\hleft(r_{*})$, 
	\begin{align}
	\hleft(r_{*}) = \frac{2}{\pi}\sqrt{\frac{3}{\sqrt{5}}} \, Q_{-\frac{1}{2}}(\tanh r_{*}).
	\label{eq:h left}
	\end{align}
	
	For $r_{*}>0$, we have 
	\begin{align}
	\hright(r_{*}) = c_{1} h_{1}(r_{*}) + c_2 h_{2}(r_{*}),
	\label{eq:h right}
	\end{align}
	with $h_{1}(r_{*})$ and $h_{2}(r_{*})$ given by Eqs. \eqref{eq:h1} and \eqref{eq:h2}. We now have to match these solutions at $r_{*}=0$ to find the constants $c_1$, $c_2$. First, continuity requires
	\begin{align}
	\hleft(r_* \to 0^{-}) = \hright(r_* \to 0^{+}).
	\label{eq:continuity condition}
	\end{align} 
	The other boundary condition is obtained by integrating Eq. \eqref{eq:h equation} inside an arbitrarily small neighborhood of $r_{*}=0$, which leads to
	\begin{align}
	\frac{d \hright}{dr_{*}}\bigg|_{r_{*}=0^{+}} - 	\frac{d \hleft}{dr_{*}}\bigg|_{r_{*}=0^{-}} = a\,  h(0).
	\label{eq:derivative condition}
	\end{align}
	This equation shows that $a$ characterizes the shape change of the nozzle at $x=0$ (see Fig. \ref{fig:extensions}).
	
	Using Eqs. \eqref{eq:continuity condition} and \eqref{eq:derivative condition}, we find
	\begin{align}
	c_1 & = -\sqrt{\frac{3}{\sqrt{5}}} \frac{\pi ^2 a}{2 \, \Gamma \left(\frac{3}{4}\right)^4}  , \\
	c_2 & =\sqrt{\frac{3}{\sqrt{5}}} \left[\frac{2}{\pi } +\frac{\pi  a}{\Gamma \left(\frac{3}{4}\right)^4}\right],
	\end{align} 
	so that 
	\begin{align}
	h(r_{*}) =& \sqrt{\frac{3}{\sqrt{5}}} \, Q_{-\frac{1}{2}}(\tanh r_{*}) \,\theta (-r_{*})   \nonumber \\
	& +\sqrt{\frac{3}{\sqrt{5}}} \left\{ -\frac{\pi ^2 a}{2 \, \Gamma \left(\frac{3}{4}\right)^4} P_{-\frac{1}{2} }(\tanh r_*) \right. \nonumber\\
	& \left.   +\left[\frac{\pi  a}{\Gamma \left(\frac{3}{4}\right)^4}+\frac{2}{\pi }\right]
	Q_{-\frac{1}{2}}(\tanh r_*)\right\} \theta(r_{*}),
	\label{eq:h extended nozzle}
	\end{align}
	where $\theta(r_{*})$ stands for the Heaviside step function. 
	The nozzle shape can then be determined by following the steps discussed in Sec. \ref{sec:The nozzle analogue to the BTZ black hole}. 
	Figure \ref{fig:extensions} shows nozzle extensions obtained for some values of the parameter $a$.
	
	We note that the diverging behavior of the cross-sectional area as $x\to \infty$ does not spoil the one-dimensional character of the motion because one can always make $A(x)$ vary as slowly as desired by suitably choosing units for $x$. As pointed out in \cite{abdalla2007perturbations}, this is equivalent to ``pulling'' the nozzle along its axis. In the present case, such a pulling means that we consider a BTZ black hole with a larger ratio $l/\sqrt{M}$.
	
	\subsection{Quasinormal modes of the extended nozzle}

	Quasinormal modes are characteristic vibrations that describe the energy loss of a system after a perturbation \cite{kokkotas1999quasi,konoplya2011quasinormal,berti2009quasinormal}. In principle, they can appear in any physical context involving open systems (not only black holes) \cite{ching1998quasinormal}.	
	Quasinormal modes in a black hole background are usually defined as mode solutions satisfying ingoing boundary conditions at the horizon ($\psi_{\omega} \sim e^{-i \omega r_{*}}$, as $ r_{*} \to -\infty$), and outgoing boundary conditions at spatial infinity ($\psi_{\omega} \sim e^{i \omega r_{*}}$, as $ r_{*} \to \infty$). This definition works perfectly well for asymptotically flat spacetimes, since the effective potential coupled to the field vanishes at infinity. However, for asymptotically curved spacetimes, the effective potential is not zero at infinity, and one cannot distinguish ingoing from outgoing modes there \cite{birmingham1997three,myung2003difference,dappiaggi2018superradiance}.
	
	As mentioned before, in contrast with the situation in asymptotically curved spacetimes, the effective potential of our extended nozzle vanishes at $r_{*} \to + \infty$. Hence one can define QNMs by the usual asymptotic behavior
	\begin{align}
	\psi_{\omega} \sim e^{-i \omega r_{*}},& \quad  \quad r_{*} \to -\infty, \label{eq:ingoing laval}\\
	\psi_{\omega} \sim e^{+i \omega r_{*}},& \quad  \quad r_{*} \to +\infty. \label{eq:outgoing laval}
	\end{align}
	The asymptotic conditions \eqref{eq:ingoing laval} and \eqref{eq:outgoing laval} completely determine the acoustic QNMs in the Laval nozzle. We will refer to modes satisfiyng Eqs.~\eqref{eq:ingoing laval} and \eqref{eq:outgoing laval} as \textit{ordinary} quasinormal modes.
	
	Quasinormal modes obeying Robin boundary conditions in the BTZ black hole were previously analyzed in \cite{dappiaggi2018superradiance}.
	In what follows, we are going to use the ordinary QNMs of acoustic waves in the nozzle to emulate QNMs of the conformal scalar field obeying RBCs in the BTZ black hole. In order to achieve this, we now calculate the former explicitly.
	
	Let us denote by $\psi_{\scriptscriptstyle <}$ and $\psi_{\scriptscriptstyle >}$ the solutions of 
	\begin{align}
	-\frac{d^{2}\psi_{\omega}(r_{*})}{dr^{2}_{*}} + V_{\mbox{{\scriptsize eff}}}(r_{*}) \psi_{\omega}(r_*) = \omega^{2} \psi_{\omega}(r_{*}),
	\label{eq:Schrodinger geral}
	\end{align}
	with effective potential given by Eq. \eqref{eq:effective potential BTZ extended}, for $r_{*}<0$ and $r_{*}>0$, respectively. The general solution of Eq. \eqref{eq:Schrodinger geral} can be expressed as a linear combination of
	\begin{align}
	\psi_{1}(r_{*}) = P_{-\frac{1}{2}}^{i \omega }(\tanh r_{*}),  \\ 
	\psi_2 (r_{*})  = Q_{-\frac{1}{2}}^{i \omega }(\tanh r_{*}),
	\end{align}
	where $P_{\nu}^{\mu }(z)$ and $Q_{\mu}^{\nu}(z)$ stand for the associated Legendre functions of the first and second kind, respectively \cite{olver2010nist}.
	
	For $r_{*}<0$, the boundary condition \eqref{eq:ingoing laval} implies 
	\begin{align}
	\psileft (r_{*}) =& \omega  \sinh (\pi  \omega ) \Gamma (-i \omega ) \psi_1( r_{*}) \nonumber\\
	& -\frac{2 i \omega}{\pi}   \cosh (\pi  \omega ) \Gamma (-i \omega )
	\psi_{2}(r_{*}). 
	\label{eq:psi left}
	\end{align}
	For $r_{*} > 0$, we have 
	\begin{align}
	\psiright (r_{*}) = c_{1} \psi_{1} (r_{*}) + c_{2} \psi_{2} (r_{*}).
	\label{eq:psi right prov}
	\end{align}
	Before considering the behavior at $r_{*}\to \infty $, we match $\psileft$ and $\psiright$ at $r_{*}=0$. Continuity requires
	\begin{align}
	\psileft(r_{*}\to 0^{-})=\psiright(r_{*}\to 0^{+}).
	\label{eq:continuidade psi}
	\end{align}
	We also require that 
	\begin{align}
	\frac{d \psiright}{dr_{*}}\bigg|_{r_{*}=0^{+}} - \frac{d \psileft}{dr_{*}}\bigg|_{r_{*}=0^{-}} = a \psi(0),
	\label{eq:derivada psi contorno}
	\end{align} 
	which is the condition obtained by integrating Eq. \eqref{eq:Schrodinger geral} inside an arbitrarily small neighborhood of $r_{*}=0$. 
	
	Solving Eqs. \eqref{eq:continuidade psi} and \eqref{eq:derivada psi contorno}, we find the constants $c_{1}$ and $c_2$ as functions of the parameter $a$. After that, we expand $\psiright$ near $r_{*} \to +\infty$,
	\begin{align}
	\psiright(r_{*}) \sim D(\omega, a) e^{-i \omega r_{*}} + E(\omega, a) e^{i \omega r_{*}}.
	\label{eq:psi right asymptotic}
	\end{align}
	The coefficients of the asymptotic expansion Eq. \eqref{eq:psi right asymptotic} are 
	\begin{align}
	D(\omega, a) = & \frac{ \pi  \text{csch}(\pi  \omega ) \Gamma (-i \omega ) }{2 i \Gamma (i \omega )} \times \nonumber\\ 
	&  \left[\frac{
		2^{2 i \omega } \pi  a}{\Gamma \left(\frac{3}{4}-\frac{i \omega
		}{2}\right)^4}+\frac{2}{\Gamma \left(\frac{1}{2}-i \omega
		\right)^2}\right],
	\label{eq:coef D freq}		
	\end{align}
	
	\begin{align}
	E(\omega, a) = & - i  \text{csch}(\pi  \omega ) \nonumber \\ 
	& - \frac{i a}{4} [\text{csch}(\pi \omega )-i]\frac{ \Gamma \left(\frac{1}{4}-\frac{i \omega }{2}\right)^2}{\Gamma
		\left(\frac{3}{4}-\frac{i \omega }{2}\right)^2}.
	\end{align}
	Hence the quasinormal frequencies of the extended Laval nozzle are given by solutions of  
	\begin{align}
	D(\omega, a) = 0.
	\label{eq:frequencias}
	\end{align}
	
	From Eq. \eqref{eq:coef D freq}, we see that ordinary quasinormal frequencies can be divided into two sets. First, since  the Gamma function has poles at negative integers, the frequencies
	\begin{align}
	\omega_{n} = - \frac{i}{2} \left( 4 n +3 \right), \quad n = 0, \, 1, \, 2 , \, 3, \,\dots
	\label{eq:dirichlet frequencies}
	\end{align}
	satisfy Eq. \eqref{eq:frequencias} for any value of $a$. 
	The second set of quasinormal frequencies is given by the solutions of 
	\begin{align}
	a = -\frac{2^{1-2 i \omega } \Gamma \left(\frac{3}{4}-\frac{i \omega
		}{2}\right)^4}{\pi  \Gamma \left(\frac{1}{2}-i \omega \right)^2}.
	\label{eq:a de omega}
	\end{align}
	
	In the following we analyze the resulting quasinormal modes for both cases, Eqs. \eqref{eq:dirichlet frequencies} and \eqref{eq:a de omega}. For convenience, we will divide the case of Eq. \eqref{eq:a de omega} in (i) $a = 0$ and (ii) $a \ne 0$.
	
	\subsubsection{Dirichlet quasinormal modes}
\begin{figure}[th!]
		\centering
		\includegraphics[width=\columnwidth]{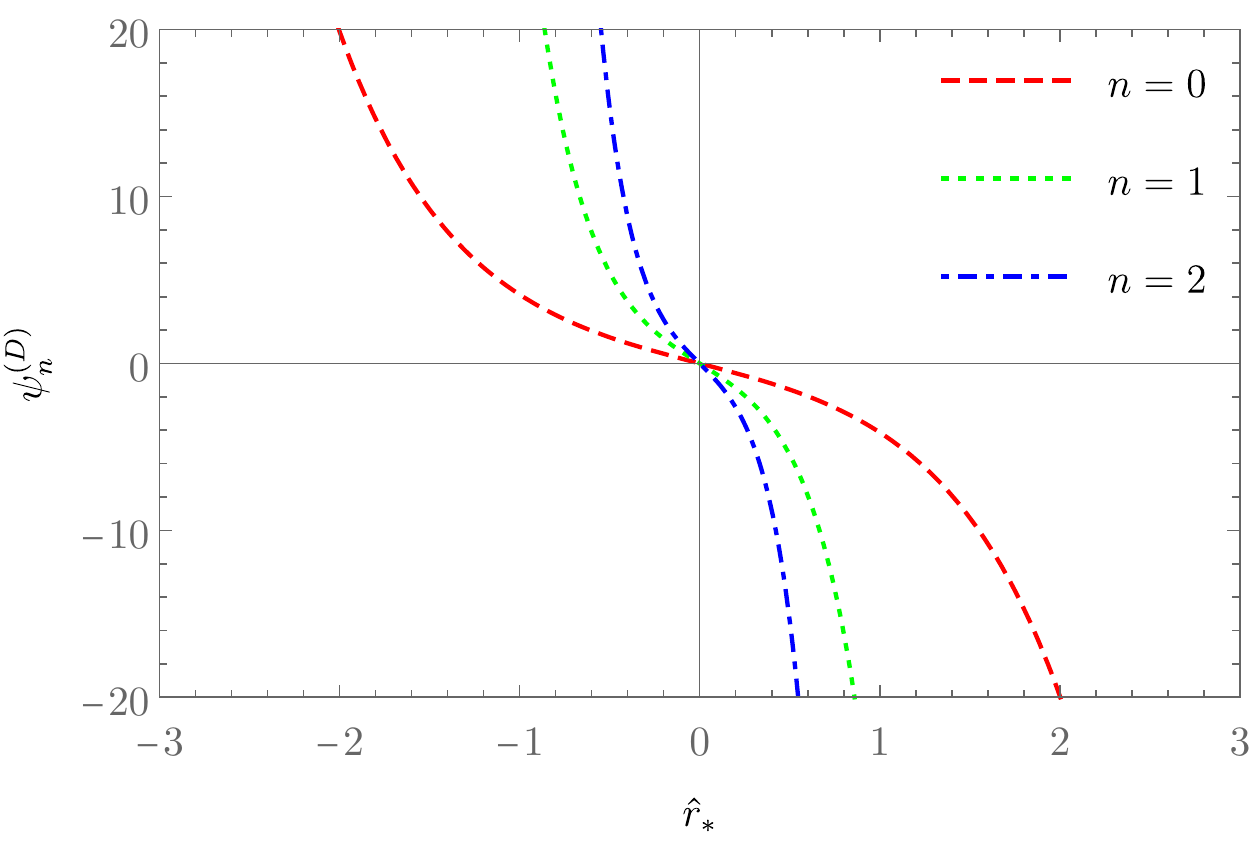}
		\caption{Spatial part of the ordinary quasinormal modes of acoustic waves in the extend nozzle as functions of the nondimensional tortoise coordinate, and frequencies given by Eq. \eqref{eq:dirichlet frequencies}. These QNMs are odd functions with respect to $\hat{r}_{*}$. For $-\infty < \hat{r}_{*} \le 0$, these mode solutions can be interpreted as QNMs of conformally coupled scalar waves obeying a Dirichlet boundary condition at spatial infinity of the BTZ spacetime. Note that these modes do not depend on the parameter $a$.
		Legend: the red dashed line represents the QNM with $n = 0$, the green dotted line represents the QNM with $n = 1$, and the blue dot-dashed line represents the QNM with $n = 2$. 
		}
		\label{fig:odd solutions}
	\end{figure}

	Let us first consider the QNMs with frequencies given by Eq. \eqref{eq:dirichlet frequencies}. Using the expressions for $\psileft$ and $\psiright$, Eqs. \eqref{eq:psi left} and \eqref{eq:psi right prov}, we find 
	\begin{align}
	\psi_n^{(D)}(r_*) =	-\Gamma \left(-2 n-\frac{1}{2}\right) P_{-\frac{1}{2}}^{2
		n+\frac{3}{2}}\left(\tanh r_*\right),
	\label{eq:dirichlet qnms}
	\end{align}
	which is defined in $-\infty < r_* < \infty$. From the transformation formula \cite{olver2010nist}  
	\begin{align}
	P_{-\frac{1}{2}}^{2n+\frac{3}{2}}\left(\tanh r_*\right)  & = \, C_n \sinh r_* \cosh^{2 n+\frac{1}{2}} r_* \times \nonumber\\
	& F\left(-n,-n;\frac{3}{2};\tanh ^2 r_*\right),
	\end{align}
	where $F$ stands for the standard hypergeometric function and
	\begin{align}
	C_n = \frac{(-1)^n 2^{2 n+\frac{5}{2}} \Gamma \left(n+\frac{3}{2}\right)}{\sqrt{\pi }
		\Gamma \left(-n-\frac{1}{2}\right)},
	\end{align}
	we see that 
	\begin{align}
	\psi_n^{(D)}(r_* =0) =0.
	\label{eq:dirichlet bc}
	\end{align}  
	Since $\psi_n^{(D)}$ is a solution of Eq. \eqref{eq:Schrodinger geral} obeying ingoing boundary conditions at the horizon $r_*=-\infty$, it follows from Eq. \eqref{eq:dirichlet bc} that, when restricted to $-\infty < r_* \le 0$, the ordinary QNM $\psi_n^{(D)}$ can be interpreted as a QNM satisfying a Dirichlet boundary condition at BTZ spatial infinity. Moreover, we note that these modes are odd functions with respect to the tortoise coordinate $r_*$, and do not depend on the value of the parameter $a$. We also mention that the frequencies given by Eq. \eqref{eq:dirichlet frequencies} are the known Dirichlet quasinormal frequencies in the BTZ background \cite{cardoso2001scalar,dappiaggi2018superradiance}. Figure \ref{fig:odd solutions} shows $\psi_n^{(D)}$ for $n = 0 , \, 1, \, 2$.
	
	\subsubsection{Neumann quasinormal modes}
\begin{figure}[th!]
		\centering
		\includegraphics[width=\columnwidth]{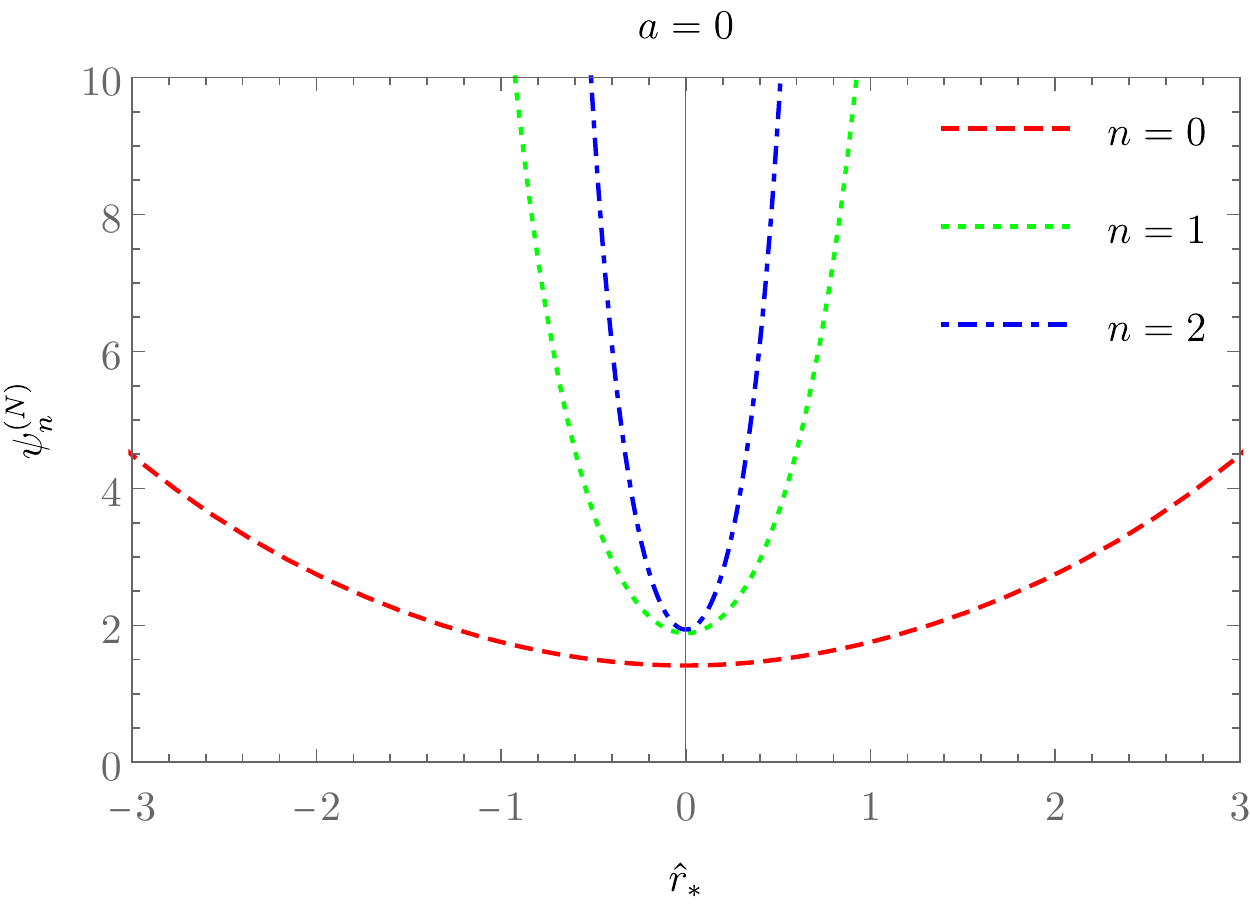}
		\caption{Spatial part of the ordinary quasinormal modes of acoustic waves in the extend nozzle as functions of the nondimensional tortoise coordinate, and frequencies given by Eq.~\eqref{eq:neumann frequencies}. These QNMs are even functions with respect to $\hat{r}_{*}$. For $-\infty < \hat{r}_{*} \le 0$, these mode solutions can be interpreted as QNMs of conformally coupled scalar waves obeying a Neumann boundary condition at spatial infinity of the BTZ spacetime, $\beta = a = 0$. 
		Legend: the red dashed line represents the QNM with $n = 0$, the green dotted line represents the QNM with $n = 1$, and the blue dot-dashed line represents the QNM with $n = 2$.
		}
		\label{fig:even solutions neumann}
	\end{figure}

	For $a = 0$, the frequencies solving Eq. \eqref{eq:a de omega} are given by 
	\begin{align}
	\omega_{k} = -\frac{i}{2}  (2 k+1), \quad k = 0, \, 1, \, 2, \, 3, \, \dots
	\end{align}
	When $k$ is odd, $k=2n+1$, these frequencies reduce to Dirichlet frequencies, Eq. \eqref{eq:dirichlet frequencies}, and the corresponding modes are given by Eq. \eqref{eq:dirichlet qnms}. On the other hand, when $k$ is even, $k=2n$, we have  
	\begin{align}
	\omega_{n} = -\frac{i}{2}  (4 n + 1).
	\label{eq:neumann frequencies}
	\end{align}
	The mode solutions in this case are given by 
	\begin{align}
	\psi_n^{(N)}(r_*) = \Gamma \left(\frac{1}{2}-2 n\right) P_{-\frac{1}{2}}^{2
		n+\frac{1}{2}}\left(\tanh r_*\right),
	\label{eq:neumann qnms}
	\end{align}
	and are defined in $-\infty < r_* < \infty$. Using the transformation formula \cite{olver2010nist}
	\begin{align}
	P_{-\frac{1}{2}}^{2 p+\frac{1}{2}}(\tanh r_*)  & = 	C_{n} \cosh ^{2
		n+\frac{1}{2}}r_* \times \nonumber\\
	& F\left(-n,-n;\frac{1}{2};\tanh ^2 r_*\right),
	\end{align}
	where
	\begin{align}
	C_{n}=\frac{ 2^{2 n+\frac{1}{2}} \Gamma \left(n+\frac{1}{2}\right)^2 }{\pi ^{3/2}},
	\end{align}
	we see that $\psi_n^{(N)}$ is an even function with respect to the coordinate $r_*$. Moreover, from the expressions above it can be shown that 
	\begin{align}
	\frac{d\psi_n^{(N)}}{dr_*}\bigg|_{r_{*}=0} =0.
	\end{align}
	Hence, when restricted to $-\infty < r_* \le 0$, the ordinary QNMs, $\psi_n^{(N)}$, correspond to QNMs satisfying a Neumann boundary condition at the spatial infinity of the BTZ black hole. We note that frequencies given by Eq. \eqref{eq:neumann frequencies} are the Neumann quasinormal frequencies for the conformally coupled scalar field in the BTZ background found in \cite{dappiaggi2018superradiance}. Figure \ref{fig:even solutions neumann} shows $\psi_n^{(N)}$ for $n = 0, \, 1, \, 2$.
	
	\subsubsection{Robin quasinormal modes}
	For $a \ne 0$, we cannot exactly solve Eq. \eqref{eq:a de omega} for $\omega$. Nevertheless, we still can show that the corresponding QNMs are even functions with respect to $r_*$. Initially we use Eq. \eqref{eq:psi left}, and substitute Eq. \eqref{eq:a de omega} into Eq. \eqref{eq:psi right prov} to find
	\begin{align}
	\frac{d \psiright}{d r_*}\bigg|_{r_{*}=0^{+}} = -\frac{d \psileft}{d r_*}\bigg|_{r_{*}=0^{-}} = -\frac{\sqrt{\pi } 2^{1+i \omega } \Gamma (1-i \omega )}{\Gamma \left(\frac{1}{4}-\frac{i \omega }{2}\right)^2} .
	\label{eq:condicao derivada origem}
	\end{align}
	Let us define $\phi(r_*)$ in $0 \le r_* <\infty$ by $\phi(r_*) = \psileft(-r_*)$. Since the effective potential is even, it follows that $\phi(r_*)$ is a solution of Eq. \eqref{eq:Schrodinger geral} in $0 \le r_* < \infty$. Moreover, we have
	\begin{align}
	\phi (0^+) &= \psileft(0^-), \label{eq:condicao 1}\\
	\frac{d \phi}{d r_*}\bigg|_{r_{*}=0^{+}} &= -\frac{d \psileft}{d r_*}\bigg|_{r_{*}=0^{-}}. \label{eq:condicao 2}
	\end{align}
	Then, by uniqueness of the solution of Eq. \eqref{eq:Schrodinger geral} obeying conditions \eqref{eq:condicao 1} and \eqref{eq:condicao 2}, we conclude that $\psileft(-r_*)=\phi(r_*)=\psiright(r_*)$. The solution in the entire interval $-\infty \le r_* <\infty$ can then be written as
	\begin{align}
	\psi_{\omega}^{(R)}(r_*) = \psileft(r_*)\theta(-r_*) + \psileft(-r_*)\theta(r_*),
	\label{eq:robin qnm}
	\end{align}
	from where it follows directly that $\psi_{\omega}^{(R)}(r_*)$ is an even function. 
	
	Another property of $\psi^{(R)}_\omega$ is found by substituting Eq. \eqref{eq:condicao derivada origem} into Eq. \eqref{eq:derivada psi contorno},  
	\begin{align}
	\beta = \frac{\left(d\psileft/dr_{*}\right)}{\psileft}\bigg|_{r_{*}=0^{-}}=-\frac{a}{2}. 
	\label{eq:robin bc nozzle}
	\end{align}
	Hence, when restricted to $-\infty < r_* \le 0$, we can interpret the ordinary QNM, $\psi_{\omega}^{(R)}$, as a QNM in the BTZ black hole satisfying a Robin boundary condition at spatial infinity with $\beta = -a /2$. 
	
	Let us analyze  Eq. \eqref{eq:robin bc nozzle} more closely. First, using Eq. \eqref{eq:a de omega} we can rewrite it as
	\begin{align}
	\beta = \frac{2^{-2 i \omega } \Gamma \left(\frac{3}{4}-\frac{i \omega
		}{2}\right)^4}{\pi  \Gamma \left(\frac{1}{2}-i \omega \right)^2}.
	\label{eq:beta prov}
	\end{align}
	Taking into account the formula \cite{olver2010nist}
	\begin{align}
	\Gamma (2 z) = \pi^{-1/2} 2^{2 z-1} \Gamma (z)	\Gamma
	\left(z+\frac{1}{2}\right)
	\end{align}
	with $z = 3/4 - i\omega/2$, we find 
	\begin{align}
	\Gamma \left(\frac{1}{2}-i \omega \right) = \pi^{-1/2} 2^{-i \omega - \frac{1}{2} }
	\Gamma \left(\frac{1}{4}-\frac{i \omega
	}{2}\right) \Gamma
	\left(\frac{3}{4}-\frac{i \omega
	}{2}\right).
	\label{eq:gamma transform}
	\end{align}
	Substituting Eq. \eqref{eq:gamma transform} into Eq. \eqref{eq:beta prov}, it follows that
	\begin{align}
	\beta = \frac{2 \Gamma \left(\frac{3}{4}-\frac{i \omega }{2}\right)^2}{\Gamma \left(\frac{1}{4}-\frac{i \omega}{2}\right)^2}, 
	\end{align}
	which agrees with the expression found in \cite{dappiaggi2018superradiance} for frequencies of quasinormal modes obeying RBCs.\footnote{The case of the $m=0$ mode of the conformally coupled scalar field in the static BTZ black hole corresponds to parameters $k=0$ and $\mu^{2}=-3/4$ in \cite{dappiaggi2018superradiance}. The parameter $\beta$ for RBCs used here relates to the parameter $\zeta$ in \cite{dappiaggi2018superradiance} by $\beta = - \cot \zeta$. \label{Dappiaggi notation}}
	
\begin{figure*}
		\centering
		\includegraphics[width=\columnwidth]{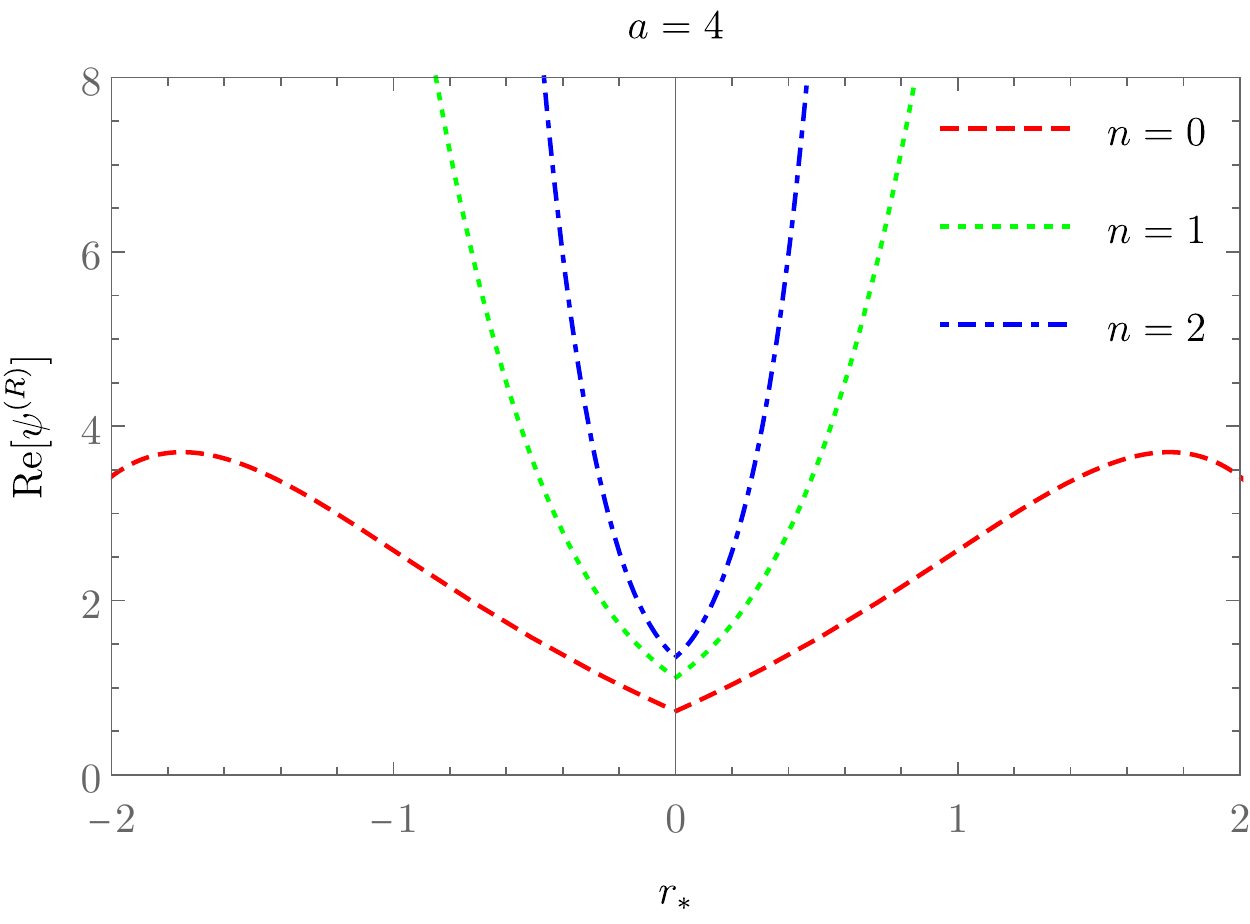}
		\includegraphics[width=\columnwidth]{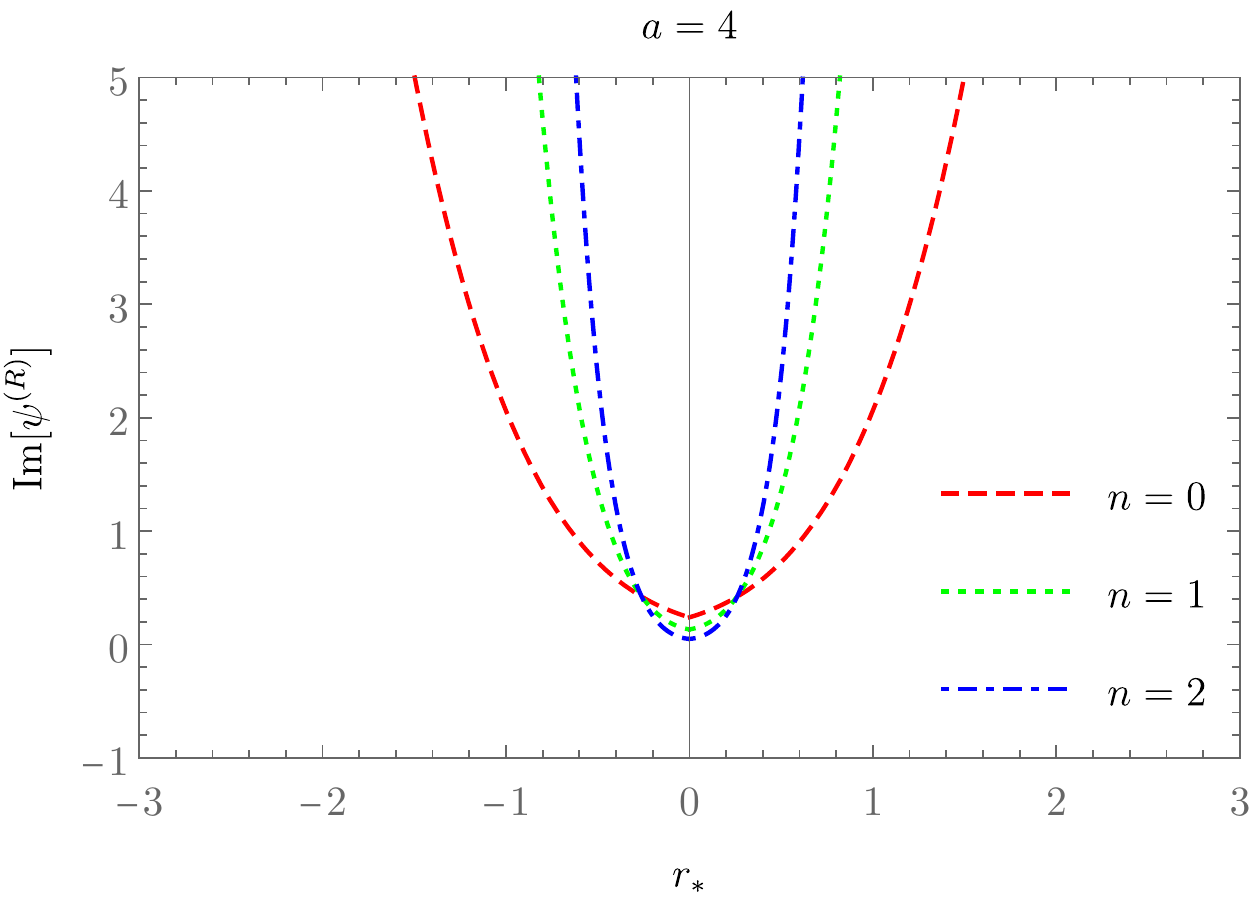}
		\caption{Spatial part of the ordinary quasinormal modes of acoustic waves in the extend nozzle as functions of the nondimensional tortoise coordinate, and with frequencies given by solutions of Eq. \eqref{eq:a de omega}. These QNMs are even functions with respect to $\hat{r}_{*}$. For $-\infty < \hat{r}_{*} \le 0$, these mode solutions can be interpreted as QNMs of conformally coupled scalar waves obeying a Robin boundary condition with $\beta = -a/2 = -2$, at spatial infinity of the BTZ spacetime. The frequencies were calculated numerically and sorted by increasing magnitude of the imaginary parts.
		Legend: The red dashed line represents the least-damped QNM, with frequency given by $\omega_0 = 0.628244 -1.21348 i $; the green dotted line represents the QNM with frequency $\omega_1 = 0.711933 - 2.69836 i $; the blue dot-dashed line represents the QNM with frequency $\omega_2 = 0.501734 - 4.54024 i$.
		}
		\label{fig:even solutions 2}
	\end{figure*}

	Figure \ref{fig:even solutions 2} shows some ordinary QNM modes of the extended nozzle with $a=4$ and frequencies given by solutions of Eq. \eqref{eq:a de omega}. The quasinormal frequencies are $\omega_0 = 0.628244 -1.21348 i $, $\omega_1 = 0.711933 - 2.69836 i $, $\omega_2 = 0.501734 - 4.54024 i$.  These ordinary QNMs correspond to QNMs satisfying a RBC in the BTZ spacetime with $\beta = -a/2 = -2$. 
	
	Summarizing the results in this section, we calculated the ordinary QNMs of acoustic waves in the extended nozzle and showed that all of them have definite parity: odd ordinary QNMs correspond to QNMs in the BTZ black hole satisfying Dirichlet boundary condition; even ordinary QNMs correspond to QNMs in the BTZ black hole satisfying Neumann or Robin boundary conditions. This provides (at least in principle) a nice way to realize Robin boundary conditions at the conformal boundary of the BTZ black hole by means of an analog model. Notice that, since for arbitrary initial data both types of QNMs (odd and even) allowed by Eq. \eqref{eq:frequencias} will be excited, in order to observe mode solutions corresponding to QNMs obeying, say, Robin or Neumann boundary conditions, one has to consider time evolution of even initial data. 
	
	Hence, we can interpret the QNMs as modes with (complex) frequencies having definite parity.  This is expected by the way the nozzle is extended. Such an extension has a  resulting even effective potential given by Eq.~(\ref{eq:effective potential BTZ extended}) so that, with  the asymptotic behavior given by Eqs.  ~(\ref{eq:ingoing laval}) and~(\ref{eq:outgoing laval}),  parity is respected.  Moreover,  Eqs.~ \eqref{eq:condicao 1} and \eqref{eq:condicao 2} show that this extension represents two images of the same  Cauchy problem with boundary conditions
	\begin{displaymath}
	\begin{array}{ll}
	\psi_\omega\sim e^{i \omega r_\ast},&r_\ast\to\infty,\\
	\frac{d\psi_\omega}{dn}(r_\ast)+\frac{a}{2}\psi_\omega(r_\ast)=0,&r_\ast\to 0, 	
 	\end{array}
	\end{displaymath}
	where $d/dn$ represents the normal derivative pointing toward  $r_{\ast}=0$.  In this way,  the quasinormal frequencies obtained in such ``extended configuration'' are precisely the ones found in the BTZ spacetime with the correspondence  $a=-2\beta$  in Eq.~\eqref{eq:robin bc}.

	Before closing this section, we mention that the lack of smoothness at the junction of the extended nozzle, resulting from the Dirac delta in the effective potential, is an idealization that could be removed, for instance, by considering a finite potential barrier in Eq.~(\ref{eq:effective potential BTZ extended}). In fact, taking a sufficiently small $\epsilon>0$, a barrier with width $2\epsilon$ and height $a/2\epsilon$ leads to a smooth nozzle with quasinormal frequencies arbitrarily close to the frequencies calculated via Eqs.~\eqref{eq:dirichlet frequencies} and \eqref{eq:a de omega}. This means that the Dirac delta in the effective potential and the resulting nonsmooth nozzle do not represent a significant limitation of our model.

	\subsection{Stability}
	\label{sub:stability}
	For black holes in asymptotically flat spacetimes, mode solutions growing exponentially in time ($\operatorname{Im}[\omega]>0$) appear as a result of energy extraction from the background spacetime by the mechanism of \textit{superradiance} \cite{brito2015superradiance}.
	
	In the case of the rotating BTZ black hole, Dappiaggi \textit{et. al.} showed that exponentially growing modes occur for a subset of RBCs \cite{dappiaggi2018superradiance}. There are two types of such modes: (i) modes corresponding to superradiant instabilities, which extract energy from the black hole; and (ii) modes arising from AdS$_{3}$ bulk instabilities \cite{dappiaggi2016hadamard}, which do not extract energy from the black hole. In both cases angular momentum is extracted from the black hole.
	
	Because our model does not account for black hole rotation, no superradiant modes occur in the quasinormal spectrum determined by Eq. \eqref{eq:frequencias}. On the other hand, since in the analog spacetime there exists exponentially growing modes that are not superradiant, we still have reason to ask if, for some value of $a$, such modes are allowed in our model.
	%These modes would correspond to instabilities coming from AdS$_3$ asymptotics of the analogue spacetime. 
	
	According to \cite{dappiaggi2018superradiance}, modes with $\operatorname{Im}[\omega]>0$ occur for RBCs with $\beta$ greater than a critical value $\beta_{c}$,
	\begin{align}
	\beta > \beta_{c},
	\end{align}
	which in our case is given by\footnote{See footnote \ref{Dappiaggi notation}.}
	\begin{align}
	\beta_{c} = \frac{2 \, \Gamma^{2} \left(3/4\right)}{\Gamma^{2} \left(1/4\right)}.
	\end{align}
	In terms of $a$, this means that unstable modes are expected to appear when 
	\begin{align}
	a < - \frac{4 \, \Gamma^{2} \left(3/4\right)}{\Gamma^{2} \left(1/4\right)} =  -\frac{2 }{\pi ^2} \Gamma^4 \left(\frac{3}{4}\right),
	\label{eq:a critico dappiaggi}
	\end{align}
	where we have used Euler's reflection formula, $\Gamma(z) \Gamma(1-z)= \pi \csc \pi z$, with $z=3/4$, to establish the last equality.
	
	From the perspective of the Laval nozzle, the expression under the square root
	in Eq. \eqref{eq:A de g} shows that the sectional area is well defined only for $g \ge 1$. One can see that this is in fact the case when $r_{*}\le0$ by noting that $\hleft$ is a strictly increasing function in the interval $-\infty< r_{*} < 0$, and has a minimum at $r_{*} \to - \infty$. Since
	\begin{align}
	\lim\limits_{r_{*}\to -\infty}	Q_{-\frac{1}{2}}(\tanh r_{*})=\frac{\pi}{2},
	\end{align}
	we see that this minimum is given by
	\begin{align}
	\lim\limits_{r_{*}\to -\infty} 	\hleft(r_{*}) = \frac{3}{\sqrt{5}} > 1.
	\end{align}
	Thus, we conclude that $g(r_{*})>1$ in $-\infty < r_{*}\le0$, for any value of $a$.
	
	For $r_{*}>0$, there are two cases to consider:
	\begin{align}
	\mbox{(i)} &  \quad  a < -\frac{2 }{\pi ^2} \Gamma^4 \left(\frac{3}{4}\right), \label{eq:case i}  \\
	\mbox{(ii)} & \quad  a\geq -\frac{2 }{\pi ^2} \Gamma^4 \left(\frac{3}{4}\right). \label{eq:case ii}
	\end{align}	
	In the Appendix  we show that for the case (i) there always exist $\tilde{r}_{*}$ such that $g(\tilde{r}_{*}) < 1$ and, hence, our model is not well defined when $a$ obeys inequality \eqref{eq:case i}. On the other hand, we show that when $a$ obeys inequality \eqref{eq:case ii} the values of $g(r_{*})$ are always greater than $1$ so that our model is well defined.	
	
	From this discussion, it follows that our model is well defined only for  
	\begin{align}
	a\ge  a_{\mbox{{\scriptsize min}}} = -\frac{2 }{\pi ^2} \Gamma^{4} \left(\frac{3}{4}\right),
	\label{eq:amin}
	\end{align}
	and, from the discussion before and including Eq. \eqref{eq:a critico dappiaggi}, we conclude that unstable mode solutions never occur in this model. This result unveils a nice feature, namely that the allowed nozzle configurations automatically reproduce only the boundary conditions that are always consistent with the stability condition in the BTZ spacetime.

	\section{summary and conclusion}\label{sec:conclusion}	
	
	We introduced an analog model for the BTZ black hole which is appropriate to analyze the QNMs resulting from Robin boundary conditions at its corresponding conformal infinity. Applying the procedure introduced in \cite{abdalla2007perturbations}, we found a Laval nozzle configuration for which acoustic waves traveling on the flowing gas mimics a conformally coupled scalar field propagating on the BTZ black hole. We found that the obtained nozzle has a finite length, and that the spatial infinity of the BTZ spacetime is mapped into one end of the nozzle. From there on, we considered nozzle extensions corresponding to effective potentials formally extending the BTZ black hole beyond its conformal infinity. In tortoise coordinates, the extended model represents two copies of the same problem,  which results in the QNMs in BTZ spacetime (with the second copy effectively extending the mode solution by parity into the region beyond the spatial infinity of the BTZ spacetime).

	After finding the ordinary QNMs in the extended nozzle, we showed that these modes can be used to simulate  QNMs in the BTZ spacetime satisfying Dirichlet, Neumann and Robin boundary conditions at its conformal boundary. Finally, we showed that, for $m=0$, the range of the parameters for which our model is well defined corresponds precisely to the range of Robin boundary conditions that allow only stable QNMs. 
	
	Although we have restricted ourselves to mode solutions with zero angular momentum ($m=0$), the case of $m\ne 0$ can be treated in a similar fashion if we take the black hole mass as $M=m^{2}$, which turns the effective potential of Eq.~\eqref{eq:effective potential BTZ nondimensional} into $\hat{V}(\hat{r}_{*}) = (5/4) \sech^{2} \hat{r}_*$. As in the case of $m=0$, the nozzle has a finite length and one can emulate RBCs by extending it, with the addition of a delta function term $a \delta(r_*)$ to the potential. Our other results still hold in this case; namely, the odd (even) ordinary QNMs correspond to QNMs obeying Dirichlet (Neumann or Robin) boundary conditions at the BTZ conformal infinity; and our model is well defined for $a \ge a_{\mbox{{\scriptsize min}}}$, for a certain $a_{\mbox{{\scriptsize min}}}$.
	The minimum value $a_{\mbox{{\scriptsize min}}}$ still constrains the range of allowed boundary conditions to an interval $-\infty < \beta \le \beta_{\mbox{{\scriptsize max}}}$, but now $\beta_{\mbox{{\scriptsize max}}}$ is smaller than the corresponding critical value $\beta_{c}$ (and therefore the allowed nozzle configurations again reproduce only the boundary conditions that are consistent with the stability condition in the BTZ spacetime).

	\acknowledgments
	The authors thank M. Richartz for insightful discussions. C. C. O. acknowledges support from the Conselho Nacional de Desenvolvimento Cient\'{i}fico e Tecnol\'{o}gico (CNPq, Brazil), Grant No. 142529/2018-4. R. A. M. was partially supported by Conselho Nacional de Desenvolvimento Científico e Tecnológico under Grant No. 310403/2019-7. J.P.M.P. was partially supported by Conselho Nacional de Desenvolvimento Científico e Tecnológico (CNPq, Brazil) under Grant No.  311443/2021-4.

	\appendix 
	\section*{Appendix: Behavior of $g(r_{*})$ for $a$ obeying inequalities $(91)$ and $(92)$}
	\label{appendix 1}
	Suppose that $a$ satisfies Eq. \eqref{eq:case i}. Noting that 
	\begin{align}
	\lim\limits_{r_{*}\to +\infty}	Q_{-\frac{1}{2}}(\tanh r_{*})=+\infty,
	\label{eq:Q infinito}
	\end{align}
	it follows that
	\begin{align}
	\lim\limits_{r_{*}\to +\infty}	\hright(r_*)=-\infty.
	\end{align}
	Hence we conclude that there exists $\tilde{r}_{*}$ such that $\hright(\tilde{r}_{*})=0$. For this $\tilde{r}_{*}$, we have $g(\tilde{r}_{*})<1$, and our model is not well defined for case (i).
	
	Let us now analyze case (ii), given by inequality \eqref{eq:case ii}. From Eq. \eqref{eq:h extended nozzle}, we have that 
	
	\begin{align}
	\hright (r_*) & =	\sqrt{\frac{3}{\sqrt{5}}} \left\{ \frac{2}{\pi }	Q_{-\frac{1}{2}}(\tanh r_*)   \right. \nonumber\\
	& \left.   +\frac{\pi  a}{\Gamma \left(\frac{3}{4}\right)^4} \left[ Q_{-\frac{1}{2}}(\tanh r_*)  -\frac{\pi }{2} P_{-\frac{1}{2} }(\tanh r_*) \right] \right\} \nonumber \\
	& \ge \sqrt{\frac{3}{\sqrt{5}}}  P_{-\frac{1}{2} }(\tanh r_*),
	\label{eq:prov 1}
	\end{align}
	where we have used the condition \eqref{eq:case ii} and the fact that
	\begin{align}
	Q_{-\frac{1}{2}}(\tanh r_*) \ge \frac{\pi}{2} P_{-\frac{1}{2} }(\tanh r_*)
	\label{eq:-k +Q}
	\end{align}
	for $0 \le r_* < \infty$. Inequality \eqref{eq:-k +Q} follows from the relation~\cite{olver2010nist}
	\begin{align}
	Q_{-\frac{1}{2}}(\tanh r_{*})  = \frac{\pi}{2} P_{-\frac{1}{2} }(- \tanh r_*)
	\end{align}	
	and the fact that $Q_{-1/2}$ is an increasing function in $-\infty < r_{*} < \infty$.
	
	Since 
	\begin{align}
	P_{-\frac{1}{2} }(\tanh r_*) &\ge 1, \qquad  0 \le r_* < \infty,
	\label{eq:minimo k}
	\end{align}
	the result \eqref{eq:prov 1} implies  
	\begin{align}
	\hright(r_{*}) \ge \sqrt{\frac{3}{\sqrt{5}}} > 1.
	\end{align}
	Thus it follows from Eq. \eqref{eq:g de h} that $g(r_{*})> 1$ whenever the constraint \eqref{eq:case ii} is fulfilled, which means that our model is well defined for $a$ in case (ii).

\end{document}